\documentclass{article}
\font\sqi=cmssq8
\def\DR{\rm I\kern-1.45pt\rm R}
\def\DC{\kern2pt {\hbox{\sqi I}}\kern-4.2pt\rm C}

\newcommand{\nn}{\nonumber\\}

\newcommand{\cZ}{{\cal Z}}

\newcommand{\bz}{{\bar z}}

\newcommand{\cbZ}{\overline{\cal Z}}

\newcommand{\bF}{{\overline F}}
\newcommand{\bQ}{{\overline Q}}
\newcommand{\bTheta}{{\overline\Theta}}

\newcommand{\bpsi}{{\bar\psi}}

\newcommand{\ba}{\begin{array}}
\newcommand{\ea}{\end{array}}
\newcommand{\be}{\begin{equation}}
\newcommand{\ee}{\end{equation}}
\newcommand{\bea}{\begin{eqnarray}}
\newcommand{\eea}{\end{eqnarray}}
\newcommand{\bi}{\begin{itemize}}
\newcommand{\ei}{\end{itemize}}

\newcommand {\bD}{\overline{D}}

\usepackage{amscd,amsmath,amssymb}

\begin{document}

\begin{center}
{\large\bf A surprise in mechanics with nonlinear chiral supermultiplet}\\
\vspace{0.3cm}
{\large Stefano Bellucci${}^{a}$ and
 Armen Nersessian${}^{b}$}
\end{center}
{\it ${}^a$INFN-Laboratori Nazionali di Frascati, Via E. Fermi 40,
00044 Frascati, Italy}\\
{\it ${}^b$\it Artsakh State University, Stepanakert\& Yerevan Physics Institute,
Yerevan,
Armenia}\\
{\sl E-mails: bellucci@lnf.infn.it, arnerses@yerphi.am
 }

\begin{abstract}\noindent
We show that the nonlinear chiral supermultiplet allows one to construct, over given two-dimensional bosonic mechanics,
the   family of
two-dimensional ${\cal N}=4$ supersymmetric mechanics parameterized with the  holomorphic function $\lambda (z)$.
We show, that this family includes, as a particular case, the
 ${\cal N}=4$ superextensions of  two-dimensional  mechanics with magnetic fields,
which have factorizable Schroedinger equations.
\end{abstract}
\subsubsection*{Introduction}
\noindent
Since its discovery \cite{witten}  supersymmetric mechanics attracts  much interest
 as a  convenient toy model for the study of dynamical consequences of supersymmetry.
It is  also a convenient  object for  developing the supersymmetry technique, particularly
 for the construction of  supersymmetric models within the superfield approach.
However, even in the latter case supersymmetric mechanics was found to have some specific properties,
which have no analogs in dimensions higher than one.
For instance, in \cite{ikl1} it was found,
 that in  ${\cal N}=4, d=1$ supersymmetry,  besides the  five off-shell linear
finite supermultiplets \cite{GR} and  the one-dimensional analog of
the ${\cal N}=2, d=4$ nonlinear multiplet \cite{dw}, there exists some new nonlinear supermultiplet,
 (called in \cite{ikl1} nonlinear chiral supermultiplet)
which  seems to have  no known higher-dimensional analogs. It  includes, as a limiting case,
 the standard chiral supermultiplet  and has the same components as the latter.
Let us recall that the standard (linear) chiral supermultiplet corresponds
to the complex superfield parameterizing the two-dimensional plane
$\DR^2=\DC^1$. Opposite to that case,  the nonlinear chiral supermultiplet corresponds to
the complex superfield parameterizing the two-dimensional sphere
(complex projective plane) $S^2=\DC P^1=SU(2)/U(1)$, but it has the same component content, as the linear one. Consequently,
the standard chirality condition is modified as follows:
\be\label{eq4} D_i \cZ = - \alpha\cZ \bD_i \cZ ,\quad \bD_i \cbZ =
\alpha\cbZ D_i \cbZ \;,\qquad  \alpha={\rm const}.\ee
In Ref.\cite{1}, by the use of the nonlinear  chiral supermultiplet,
  the  model of two-dimensional ${\cal N}=4$ supersymmetric
mechanics has been suggested, with the  following superfield action:
\be\label{action1}
S = \int\! dt d^2\theta d^2 \bar\theta\; K(\cZ,\cbZ)
+ \int\! dt d^2 \bar\theta\; F(\cZ) + \int\! dt
    d^2\theta\; \bF (\cbZ) \;.
\ee
Here $K(\cZ,\cbZ)$ is an arbitrary real function playing the role of K\"ahler potential of the metric, while $F (\cZ)$ and
$\bF(\cbZ)$ are arbitrary holomorphic  and antiholomorphic functions.
Some interesting features of the model were observed there, e.g.
 the possibility to incorporate
a magnetic field preserving the supersymmetry of the system.
 It was shown that  this system includes, as  a particular case,  the ${\cal N}=4$
supersymmetric  Landau problem on the sphere. Later on the nonlinear chiral multiplet has been used
for the construction of ${\cal N}=8$ supersymmetric mechanics \cite{n8nonlin}, as well as obtained by the reduction of the
linear supermultiplet with four bosonic and four fermionic degrees of freedom \cite{root}.

In the present  note we show that ${\cal N}=4$ supersymmetric mechanics with nonlinear chiral multiplet possesses a
quite surprising property.

When we construct the supersymmetric mechanics with linear chiral multiplet,
the arbitrariness  of the construction is in the choice of
K\"ahler potential $K$ and superpotential $F(z)$ only, and these functions define the  underlying bosonic configuration.
The  extension to a supersymmetric system is unique \cite{ber}.
On the contrary, when dealing with  nonlinear  chiral supermultiplet,
we have the  freedom in the supersymmetric extension of the given bosonic system,
encoded in the choice of the holomorphic function $\lambda(z)$.
When the underlying  bosonic system is of the sigma-model type, the  function  $\lambda(z)$ remains arbitrary.
 Otherwise it is related with the given potential and magnetic field as follows:
\be
U(z,\bar z) =\frac{F'(z){\bF}'(\bz)}{(1+\lambda{\bar\lambda})^2 g}\;,\qquad
B=\frac{{\bar \lambda}'(\bz) F'(z)+
\lambda'(z) { \bF}'({\bz})}{\left(1+\lambda{\bar\lambda}\right)^2g}\;.
\label{UB}\ee
Here $gdz d\bz$ defines the K\"ahler metric of the underlying bosonic space, $U$ is a potential of the underlying bosonic system,
 and $B$ is the magnitude of the magnetic field.

An interesting feature of supersymmetric (quantum) mechanics is the application to integrable systems of quantum mechanics.
Initially, it was found that supersymmetric quantum mechanics could be naturally related with the  factorisation method of the
 solution of one-dimensional Schroedinger equations, yielding the algebraic approach
to the construction of the spectra of all known integrable one-dimensional quantum-mechanical systems
 \cite{Gendenshtein}. Later on, the  factorisation method, based on supersymmetric mechanics, has been applied to the specific
higher-dimensional mechanics with  spin (see, e.g. \cite{Gendenshtein2} and   \cite{sukhatme} for a review and references ).

Recently, Ferapontov and Veselov performed a systematic study of the factorisation method of  quantum mechanical
systems on curved two-dimensional surfaces in the presence of magnetic field (without any use of  the supersymmetry technique)
\cite{Ferapontov}. In particular, they found the restrictions to the admissible set of potentials and magnetic fields, which allows
for a factorisable Schroedinger equation.
We shall show that for the specific choice $\lambda=\pm F$, the system under consideration
 yields  a ${\cal N}=4$ superextension of Ferapontov-Veselov mechanics.

\subsubsection*{$\lambda(z)$-freedom}
The  action (\ref{action1}) of the ${\cal N}=4$ supersymmetric mechanics
with nonlinear chiral supermultiplet  can be represented as follows \cite{1}:
\be
S = \int dt \Biggl \{ g
    \dot{z} \dot{\bz} - i\bar\lambda(\bz){\cal F}_z \dot z
 + i\lambda(z){\cal\bF}_{\bz} \dot\bz -\frac{{\cal F}_z{\cal\bF}_{\bz}}{g}
\nn\ee
\be\label{action2}
 +   \frac{i}{4} h \left[
     \psi^i {D_t{\bpsi}_i} -({D_t{\psi}^i}) \bpsi_i +\frac{\dot\bz\lambda'\psi^2}{1+\lambda\bar\lambda}+
\frac{{\dot z}\bar\lambda'\bpsi^2}{1+\lambda\bar\lambda}
\right ]
 -
\frac{(1+\lambda\bar\lambda)}{4}\left[ \frac{h^2}{4}\left(\frac{\lambda'\bar\lambda'}{(1+\lambda\bar\lambda )h}-{\cal R}\right)\psi^2\bpsi{}^2
+{\cal F}_{z;z}\psi^2-{\cal\bF}_{\bz;\bz}\bpsi^2 \right]
    \Biggr \}.
\ee
The supercharges corresponding to the ${\cal N}=4$ supersymmetry transformations read
\be
Q^i=\Theta^i-{\bar \lambda}(\bz ) \bTheta^i,\quad \bQ_i=\bTheta_i+\lambda(z) \Theta_i \; ,\;
\label{Theta}\ee
where
\be
 \Theta^i=g\left(\dot\bz+\frac{i}{4}\bar\lambda'(\bz)\bpsi^2\right)\psi^i+i{\cal\bF}_{\bz}\bpsi^i,\quad
\bTheta_i=g\left(\dot z+\frac{i}{4}\lambda'(z)\psi^2\right)\bpsi_i+i{\cal F}_{z}\psi_i\; .
\nn\ee
Here we introduced the following notation:
\be\label{def1}
 g(z,\bz)=
 \partial\bar\partial K(z,\bz),\quad \lambda(z)=\alpha z,\quad \bar\lambda(\bz)=\alpha\bz,
\quad h(z,\bz)=(1+\lambda\bar\lambda)g,
\quad{\cal F}_z=\frac{F'(z)}{1+\lambda\bar\lambda},\quad {\cal\bF}_\bz=\frac{\bF'(\bz )}{1+\lambda\bar\lambda}
\ee
and
\be
 D\psi^i=d\psi^i+\Gamma\psi^i dz , \quad \Gamma=\partial\log h,\quad {\cal R}=-\partial\bar\partial\log h/h ,\quad
\quad {\cal F}_{z;z}=\partial{\cal F}_z-\Gamma{\cal F}_z  \; .
\label{def2}\ee
It is clear that $\Gamma$ and ${\cal R}$ define, respectively,
 the connection  and the scalar curvature  of the metric $h(z,\bz)dz d\bz$, while ${{\cal F}_{z; z}}$
is the covariant derivative of the one-form ${\cal F}_z dz$ with respect to this metric.

 The above expressions are not
covariant with respect to holomorphic transformations
$
z\to f(z), \quad \psi\to f'(z)\psi\;
$.
However, the covariance will be immediately restored, if we assume that {\sl $\lambda (z)$
is an arbitrary holomorphic function}, instead of $\lambda=\alpha z$, and $\lambda'(z)=d\lambda(z)/dz$ instead of $\lambda'=\alpha=const$.

The key observation is that when $F'=0$, the kinetic term of the underlying bosonic system does not change upon this replacement.
{\it Hence, there are   infinitely many ways to supersymmetrize
 a free particle  (i.e. when $F'(z)=0$), since in this case
 the  function $\lambda (z)$ could be any!}
This is a completely unexpected result: to our knowledge,  supersymmetry
was, in some sense, an occasional (or exceptional)  property
in  mechanical systems with fermionic and bosonic degrees of freedom, being (almost completely)
defined by the underlying bosonic configuration. Namely,  for its appearence, a  strong  correlation between
the spin interaction and the electric-magnetic one was needed.
Instead, in the present model this is not the case.

The system contains the interaction
with  a {\it nonzero magnetic field} defined by the one-form ${\cal A}_B$
\be
{\cal A}_{B}=i{\bar\lambda}(\bz){\cal F}_z dz - i\lambda (z){\cal\bF}_{\bz}d{\bz}\;,\quad
d{\cal A}_B=i\frac{{\bar \lambda}'(\bz){\cal F}_z+ \lambda'(z) {\cal \bF}_{\bz}}{(1+\lambda{\bar\lambda})^2}d\bz \wedge dz.
\label{A0}\ee
Thus,  the magnitude of the magnetic field and the bosonic potential are defined by the expression (\ref{UB}).

Hence, the appearance of the magnetic field  and  the potential yields a restriction in the freedom of  choice of the $\lambda(z)$
function, defining the coupling of the fermionic degrees of freedom; but even in this case it is not completely fixed.

\subsubsection*{Ferapontov-Veselov systems}
Let us consider the special case of the ${\cal N}=4$
supersymmetric mechanics with nonlinear chiral multiplet, when
  $\lambda(z)=\pm F(z)$ (notice, that upon the choice $\lambda=iF$ the potential term remains unchanged, but the magnetic field vanishes ).
Upon this choice one has
\be
B=\pm 2U,\quad U=\frac{F'\bF'}{(1+F\bF)^2 g}\;\; .
\ee
Such systems possess a quite important property: it was shown by Ferapontov and Veselov,
that they have a factorisable quantum Hamiltonian (Schroedinger operator) \cite{Ferapontov},
closely related with supersymmetry \cite{Gendenshtein,Gendenshtein2,sukhatme}.
Let us recall that the Schroedinger operator $\widehat{\cal H}_0$ is called factorisable, if it can be represented in the form
 $\widehat {\cal H}_0=\widehat D_1 \widehat D_2$,
where $\widehat D_{1,2}$ are first-order differential operators. In this  case the operator
$\widehat {\widetilde{\cal H}}_0=\widehat D_2 \widehat D_1$ has the same spectrum as the  former one, except,
 possibly, the state with zero energy! In the one-dimensional case  one can calculate
the complete spectra of such an operator in a purely algebraic way. Opposite to the
one-dimensional case, the two-dimensional
factorisable Schroedinger operators are quasi-exactly solvable,  while only  the operators with constant magnetic fields on the
spaces with constant curvature admit complete exact solvability \cite{Ferapontov}.
Also, in \cite{Ferapontov} it was found  that if the two-dimensional  Schroedinger equation with $B=U=0$ is
integrable on some surface,
then the Schroedinger equation with $U=\pm B/2={\cal R}_0$ (where ${\cal R}$ is the scalar curvature of the surface)
 is also integrable, and
it has the same spectrum as the  former one, except, possibly, the  zero-energy level \cite{Ferapontov}.
For the system under consideration this requirement yields the following restriction to the metrics:
\be
g(z,\bz)dzd\bz=\frac{dz d\bz}{(1+\lambda\bar\lambda )^2}.
\ee
In this case the scalar curvature of the metrics is given by the expression
\be
{\cal R}_0=\lambda'(z)\bar\lambda'(\bz )\;\; .
\ee
Notice that upon this choice of the metric,  the Lagrangian
 of the supersymmetric mechanics with nonlinear chiral multiplet looks much simpler, than with the generic one
\be
{\cal L}=  \frac{\dot z \dot\bz}{(1+\lambda\bar\lambda)^2}
-i\frac{\bar\lambda(\bz) F' \dot z -\lambda(z){\bF}' \dot\bz }{(1+\lambda\bar\lambda)}-{ F}'{\bF}
\nn\ee
\be
 +   \frac{i(\psi\dot\bpsi-\dot\psi\bpsi)}{4(1+\lambda\bar\lambda )}
+\frac{i(\dot z \lambda'\bar\lambda -\dot\bz \bar\lambda'\lambda)\psi\bpsi +
\dot z \bar\lambda'\bpsi^2+\dot{\bar z}\lambda'\psi^2}{4(1+\lambda\bar\lambda)^2}
-
\frac{{F}''\psi^2-{\bF}''\bpsi^2}{4}.
\ee
Finally, let us notice that the  factorisation method of the
Schroedinger equation is generically with ${\cal N}=2$ supersymmetry, describing  particles
with spin $1/2$. In the presented case we  arrive, after quantization,
to a spin $1$ system.  It seems  clear that factorising our supersymmetric  (quantum) Hamiltonian,
 we shall arrive to the pair of  isospectral  Hamiltonians for  spin $1/2$ systems.
\subsubsection*{Conclusion}
 We have shown that the $N=4$ supersymmetric mechanics with the nonlinear chiral
supermultiplet, constructed on $S^2=SU(2)/U(1)$ qualitatively differs
from other  supersymmetric mechanics models constructed within the superfield approach.
The difference  insists in the  wide freedom in the supersymmetrization ways of the given bosonic system,
encoded in the choice of the holomorphic function $\lambda(z)$.
When the underlying bosonic system has no interaction with the external field this function remains arbitrary.
Otherwise, it is restricted by the given form of potential and magnetic field. A particular choice of the $\lambda(z)$-function
allows one to include in the class of ${\cal N}=4$ supersymmetrizable mechanics  the two-dimensional systems with
factorisable Schroedinger equation, analyzed by Ferapontov and Veselov.
We believe that this simple example may drastically change the common
intuitive impression about the rigidity of the supersymmetrization procedure.

Let us notice that by the use of chiral supermultiplet one can construct the ${\cal N}=8$ supersymmetric mechanics as well
\cite{ikl1,n81}. The use of  linear chiral multiplet yields the supersymmetric mechanics on special K\"ahler manifolds, with
a strong restriction on the admissible set of potentials \cite{n81}.
The  ${\cal N}=8$ supersymmetric mechanics with nonlinear chiral supermultiplet \cite{n8nonlin} has a metrics with
the deformed condition ensuring that the space be a special K\"ahler one.
We are sure that, similarly to the above consideration, also in this case one can restore  the $\lambda(z)$ freedom, may be
with some additional restriction, as well as  clarify the origin of that deformation.
Also, notice that in the recent paper \cite{root} the  linear and nonlinear  chiral supermultiplets
were obtained by the reduction of  the linear supermultiplet with
four bosonic and four fermionic degrees of freedom \cite{ikl1,Ivanov}. However, it is still unclear, how $\lambda(z)$-freedom could
be explained in this picture.\\

{\large Acknowledgements.}
The authors thank Sergey Krivonos and Emanuele Orazi for  valuable discussions and their interest in this work.
This research was partially supported by the European
Community's Marie Curie Research Training Network under contract
MRTN-CT-2004-005104 Forces Universe.

\end{document}